# The Most Severe Test for Hydrophobicity Scales: Two Proteins with 88% Sequence Identity but Different Structure and Function

Alexander E. Kister* and James C. Phillips[§]

*Univ Med & Dent New Jersey, Dept Hlth Informat, Sch Hlth Related Profess, Newark, NJ  07107

§ Dept. of Physics and Astronomy, Rutgers University, Piscataway, N. J., 08854

**Protein-protein interactions (protein functionalities) are mediated by water, which compacts individual proteins and promotes close and temporarily stable large-area protein-protein interfaces**.  **In their classic paper Kyte and Doolittle (KD) concluded that the "simplicity and graphic nature of hydrophobicity scales make them very useful tools for the evaluation of protein structures".  In practice, however, attempts to develop hydrophobicity scales (for example, compatible with classical force fields (CFF) in calculating the energetics of protein folding) have encountered many difficulties.  Here we suggest an entirely different approach, based on the idea that proteins are self-organized networks, subject to finite-scale criticality (like some network glasses).  We test this proposal against two small proteins that are delicately balanced between α and α/β structures, with different functions encoded with only 12% of their amino acids.  This example explains why protein structure prediction is so challenging, and it provides a severe test for the accuracy and content of hydrophobicity scales.  The new method confirms KD's evaluation, and at the same time suggests that protein structure, dynamics and function can be best discussed without using CFF.**

Per Bak originated the idea that criticality, previously considered to be a property of continuum models, might also be useful in describing the properties of networks, providing that they are self-organized[1].  He suggested that this idea could be applied to evolution, 1/f noise, and punctuated equilibrium.  The idea has since been applied in many contexts, but in practice primarily to seismic phenomena and more recently neural networks[2].  Here we test this idea against protein structure and function, for which there is a large and rapidly growing data base.

The idea has more than one source: quite early, one of us suggested that some network glasses could be self-organized, and that this would optimize their ability to fill space with a constrained and compacted ("folded"), yet unstressed (easily deformed), network[3]. The relevance of these ideas to proteins became much greater after the discovery of the reversibility window in the phase diagrams of both chalcogenide and oxide network glasses, as well as solid electrolytes[4,5]; this discovery solves the otherwise inexplicable problem of entanglement of exponentially complex, deformed networks, which overhung all such networks (network glasses as much as proteins).

How amino acid sequence determines protein structure and ultimately protein function is perhaps the most fundamental unresolved question in biology. The degenerate nature of folding information, the vast number of conformations available to a 300 amino acid (aa) polypeptide chain, and the low stability of most natural proteins ($\Delta G_{unfolding}$ between 5 and 15 kcal/mol), all present challenges to understanding the folding code. The scope of the difficulties can be appreciated when one realizes that a far simpler problem, predicting whether simple binary octet compounds $A^N B^{8-N}$ (no lone pairs) crystallize as six-fold coordinated salts with ionic bonding (NaCl, CaO), or as four-fold coordinated semiconductors with covalent bonding (Si, GaAs, ZnSe), was not solved until the late 1960's. Strikingly, the solution does not involve elaborate first-principles quantum calculations (which cannot predict sufficiently small energy differences between ionic and covalent structures, even today), but the construction of an absolute (100% accurate) scale of fractional ionicity of these simple chemical bonds, based only on their observed valence electron density and electronic dielectric constants, which measure average energy gaps in response to large-scale electric dipole perturbations[6].

Although nearly everyone agrees that a truly accurate hydrophobicity scale could be of great value, both conceptually and practically, many have been discouraged by the long history of failed attempts to construct such scales[7]. The heart of the problem, discussed at length elsewhere[8], lies in coupling such attempts to the energetics of short-range classical force fields (CFF) and/or individual amino acid solvent accessible surface areas (SASA). Instead, in analogy with fractional ionicity, what is needed is a *dimensionless* long-range hydrophobicity scale; a way to construct such a scale was recently discovered[9], based on exponents[8] obtained from the scaling of SASA and self-organized criticality. An important feature of the discovery is that it is fully compatible with the finite-size cutoff implicit in the lengths of protein secondary structures[8], which are generally no longer than 30-40 aa's, and with the doubly (water + protein) percolative properties of the reversibility window of polymer blends and network glasses[10].

The application chosen here to demonstrate the validity of the Zebende *dimensionless* hydrophobicity scale appears to provide the most severe test possible. Protein G, a multidomain cell wall protein, contains two types of domains that bind to serum proteins in blood: $G_A$ domains of 47 structured amino acids that bind to HSA, and $G_B$ domains of 56 structured amino acids that bind to the constant (Fc) region of IgG[11]. The $G_A$ and $G_B$ domains share no significant sequence similarity and have different folds, 3-α and ββαββ, respectively. The $G_A$ and $G_B$ crystal structures have been studied complexed with their hosts, and the (contact) binding sites clearly identified. The $G_A$ domains were augmented to 56 aa's; thermal functional stabilities and structures of many $G_A$ and $G_B$ mutants were studied, culminating in two proteins with 88% sequence identity, which preserved both their original folds and binding functions. According to the authors, their results not only pose a problem for traditional computational approaches to protein folding based on CFF, but they also suggest that "the three characteristics of proteins that make the folding problem difficult (large conformational space, degenerate folding code, and small ∆G) may enable facile evolution of new folds and functions". We believe that all of these factors are included in the concept of finite-size self-organized criticality occurring in the protein reversibility window, and that this underlies the widespread successes of the Zebende *dimensionless* hydrophobicity scale reported elsewhere[8] and here.

**Results and Discussion**

Here we test three hydrophobicity scales[8]: the classic KD scale based on "an amalgam of experimental observations [on isolated amino acids] derived from the literature", based chiefly on transference energies of individual aa's from water to organic solvents[12]; a water-only scale (with dimensions of length), based on mean buried depths[13], and the Zebende *dimensionless* hydrophobicity scale[9]. We *never* use isolated (short-range) hydrophobicity values ψ1, which correspond to conventional (non-contextual) SASA; instead, we always smooth the inputs by averaging over N sites as $\langle\Psi N\rangle$. There are two natural choices for N. The obvious one is to average over a structural element (such as a helix or strand), important for stability. Another possibility, useful for studying binding on any length scale, is to average over N = 3 sequential residues.

A remark about $\langle\Psi 3\rangle$ is in order here. At first sight this contextual average appears to involve nothing more than rectangular smoothing to reduce noise, but it has a much deeper justification. First, as regards self-similarity, the Zebende exponents begin with N = 1 in 2N+1 residue groups, thus $\langle\Psi 3\rangle$ is a natural choice. Moreover, the i, j = i±4 α helical periodicity is such that $\langle\Psi 3\rangle(i)$ never overlaps $\langle\Psi 3\rangle(j)$; thus the moving average contains hidden spiral information. It is probably for these two reasons, as well as the general tendency towards hydroneutrality, that quite consistently our applications have shown interesting $\langle\Psi 3\rangle$ patterns.

**Helix/Strand Stability Patterns**

We begin with the "easy" question of the relative stabilities of $G_A$, $G_B$ and their mutants, and focus on the $\langle\Psi N\rangle$ averages of the rigid $G_A$ $\alpha_1\alpha_2\alpha_3$ and $G_B$ $\beta\beta\alpha\beta\beta$ helical elements[8]. For graphical convenience, we construct for $G_A$ the model five-component secondary sequence $\alpha_1[(\alpha_1 + \alpha_2)/2]\alpha_2[(\alpha_2 +\alpha_3)/2]\alpha_3$, and compare this to the five-component $G_B$ secondary sequence $\beta_1\beta_2\alpha\beta_3\beta_4$. The results for the three hydrophobicity scales are shown in Fig. 1: what is one to make of them? The answer is, quite a lot. As we shall see below, the active sites lie near the middle of each protein, and this makes the large-scale symmetrical patterns shown for the Zebende scale quite attractive. At first glance, the results shown for the Pintar scale look similarly symmetrical, but there is an important difference: overall, they are more hydrophilic, while the Zebende results are centered near hydroneutrality at 0.155. (The ordinates in all the figures are multiplied by $10^3$; most proteins are delicately balanced near hydroneutrality.) The KD scale gives similar results for GA (all helices), but its results for $G_B$, although still symmetrical, look ragged. Finally, and this is a common feature for all scales, but it is clearest for the Zebende scale, the $G_B$ pattern's symmetry mirrors the $G_A$'s: the former has a valley near its center, while the latter has a hill. This is gratifying, as it hints at a correlation between differences in large-scale hydrophobicity patterns and the different binding functions of $G_A$ and $G_B$.

In Fig. 2 we see the effects of mutation to 88% sequence identity for the Zebende scale. The $G_A$ curve is flattened (as one would have expected, it moves towards the $G_B$ curve). However, the behavior of the latter is so unexpected that it was checked several times. Strand 3 (position 4 in the figure) not only moves towards $G_A$, but it actually overshoots and becomes a secondary peak for $G_B$ (different from $G_A$). One would think that this extra oscillation in $\langle\Psi N\rangle(G_B)$ is unfavorable, and Fig. 3, which shows the effects of mutation on stability (ΔG in kcal/mol) and overall hydrophobicity $\langle\Psi 56\rangle$ (measured relative to a baseline of 1440, thus $\langle\Psi 56\rangle$ = 5.0 in the figure corresponds to a "true" value of -0.1490 for the Zebende exponent), confirms this. The stability of $G_A$ drops by 1/3 on mutation, but that of $G_B$ drops by more than a factor of 3. The correlation of large-scale hydrophobic averages $\langle\Psi N\rangle$ with the internal structure and overall stability might appear to be coincidental, but we have already observed many such "coincidences" in repeat proteins (large, and with different structures and functions)[8]. Instead, we regard this success as reflecting the close connections between structure and function mentioned in [11], and which we attribute to finite-scale self-organized criticality occurring in the protein reversibility window.

For completeness Table I shows the $\langle\Psi 56\rangle$ values for the three scales. Knowing that the Zebende scale gives the best results elsewhere, as well as its having a strong theoretical foundation, we can see from the Table that the other two scales are unreliable, even when averaged over a large number of aa's.

**Functional Binding Patterns**

It is customary to discuss protein binding to a substrate in terms of individual aa near-contacts observed crystallographically *ex vitro*, but it is important to remember that the *in vivo* interactions occur *in vitro*, and with thermal averaging over many configurations not represented crystallographically. Here we focus on chemical trends in hydrophobicities of $G_A$ and $G_B$ in the ranges of contacts that define crystallographic binding. We define two such ranges, a simpler narrow one (from 28 to 36, comprising all the IgG binding sites in $G_B$ on the central helix (E27,

K28, K31, and N35), and many of the binding sites of $G_A$, and a wider one, from 21 to 46, which includes all the nominal (contact) binding sites of both proteins, and both the central helix and strand 3 of $G_B$.

Fig. 4 shows the results of the three scales in the narrow region. In order for the mutations to 88% sequence identity to be able to preserve function for both $G_A$ and $G_B$, it is necessary that these two curves follow each other closely in the binding region; otherwise, during mutation, one or both proteins functionalities would almost surely deteriorate unacceptably. The Figure shows that this close similarity appears with the Zebende self-similar scale, but it does not appear in either the KD amalgam scale, or the Pintar buried scale; although the latter are similar to each other, they are merely incompatible with experiment in the same way. On the one hand, this might reflect the non-scaling length dependence of the KD and Pintar (individual aa) scales, but on the other hand it shows why experimentalists have historically had little confidence in hydrophobicity scales. The way the sequence convergence actually took place experimentally is shown in Fig. 2.

When we look at the patterns in the wide region, Fig. 5, our first impression is that the three patterns look rather similar. This is as it should be: after all, the three scales all describe the same phenomenon of hydrophobic compaction. Unfortunately, the largest differences occur at the center of the narrow binding region (Fig. 4), and this just happens to be the region of greatest functional importance. The moral of this story is simple: proteins are complex, and a "good" theory is not good enough; only a great theory, one that is extremely accurate, perhaps even absolute, is sufficient. From the point of view of scaling theory, it is striking that the three scales (which scale as $L^\varepsilon$, with $\varepsilon = 0,1,2$ for Zebende,Pintar,KD) can give such similar results qualitatively, while the exponents are so different, yet this similarity disappears at the center of the binding ranges, reinforcing the view of ref. [11] and here concerning the close connections between sequence and function (presumably evolutionary in origin). At this point, the reader should also recall that the $\langle\Psi N\rangle$ rigidity patterns of $G_A$ and $G_B$ (discussed earlier) mirror each other, and that apparently explains the differences in function.

By now the reader may have gained the impression that the hydrophobicity patterns of $G_A$ and $G_B$ are similar throughout their entire 56 aa sequences, but this is not so. In Fig. 6 the patterns are compared near the N and C terminals, and it is seen that there is little similarity; the similarities are confined to positions 22-39 (approximately 1/3 of the entire sequence, which contains 4 of the 7 positions that were not mutated). This is the kind of intriguing problem that arises with hydroanalysis: is this pattern common to all proteins of this or even a larger family? Here we will, however, stop and leave this question to another time.

In conclusion, it is tempting to suppose that as time passes and we learn more about accurate hydrophobicity patterns and their relations to stability and function, Kyte and Doolittle's original hope, that the "simplicity and graphic nature of hydrophobicity scales make them very useful tools for the evaluation of protein structures", will be realized. Should that happen, the entire question of protein folding[14] may fade into the background, and be replaced by a genuine super-homology theory that relates sequence, structure, stability and function[8].

*Postscript.* While our approach apparently yields useful results on large-scale stabilities, it is not designed to be reliable for the effects of individual mutations on stability. In that limit one can expect only qualitative results, as the free energy changes are known to be contextual, that is, scattered and not determined by the individual mutation alone. Given that qualification, a reasonably accurate helix propensity scale based on experimental studies of peptides and proteins is available[15]. The case of alpha-helix stabilization by alanine relative to glycine[16] is especially interesting, as these two aa's have practically the same Zebende hydrophobicities. In internal helical positions, A is regarded as the most stabilizing residue, whereas G, after proline, is the most destabilizing. The studies showed a significant effect when the mutation was at the center of a helix, but at the edge of a helix (a "C1 position"), no effect was observed[16].

In $G_{A,B}$ position 24 is one of the seven unmutated positions in G88, occupied by G,A. Position 24 is just outside helix 1 in $G_A$, and just inside the central helix in $G_B$. Experiment showed[17] that mutating either site destabilizes by about 0.5 kcal/mol, which appears to be the limit of accuracy of helix propensity studies of individual mutations.

|      | Z     | P     | KD    |
|------|-------|-------|-------|
| GA   | 146.2 | 129.7 | 159.6 |
| GA88 | 150.2 | 128.5 | 165.6 |
| GB88 | 145.2 | 128.9 | 155.6 |
| GB   | 144.2 | 127.4 | 152.5 |

Table I. Values of $\langle \Psi 56 \rangle$ for the three scales discussed here, for wild and mutated proteins.

**Figure Captions**

Fig. 1. Hydrophobicity averages for the three scales discussed here, over the strands and helices of $G_A$ and $G_B$, as labeled in the text.

Fig. 2. Comparison of wild and mutated proteins near the center of the binding region, using the scale of [9]. The red and green curves are nearly identical.

Fig. 3. Stability and average hydrophobicity <Ψ56> as evolved from wild to final mutations of [11].

Fig. 4. Comparison of sequential hydrophobicities in the central binding region for the three scales discussed here.

Fig. 5. Comparison of sequential hydrophobicities in the wide binding region for the three scales discussed here.

Fig. 6. Comparison of sequential hydrophobicities near the C and N terminals (outside the central binding region) for the three scales discussed here.

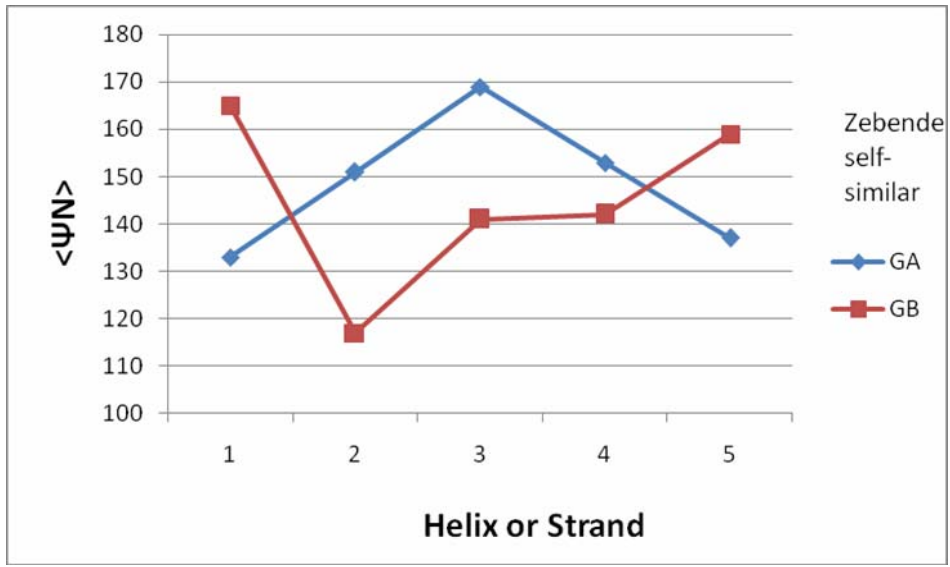

Fig. 1 (a)

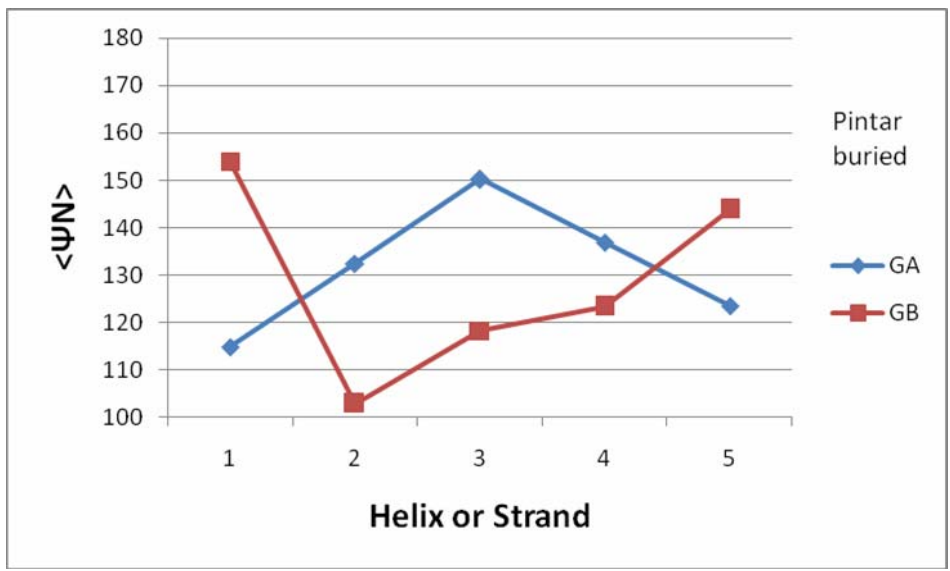

Fig. 1(b)

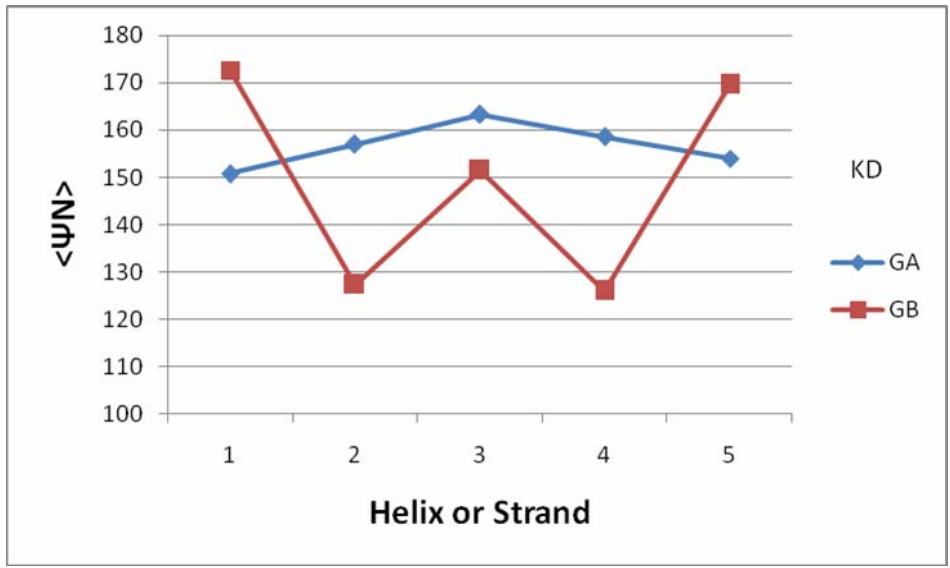

Fig 1(c)

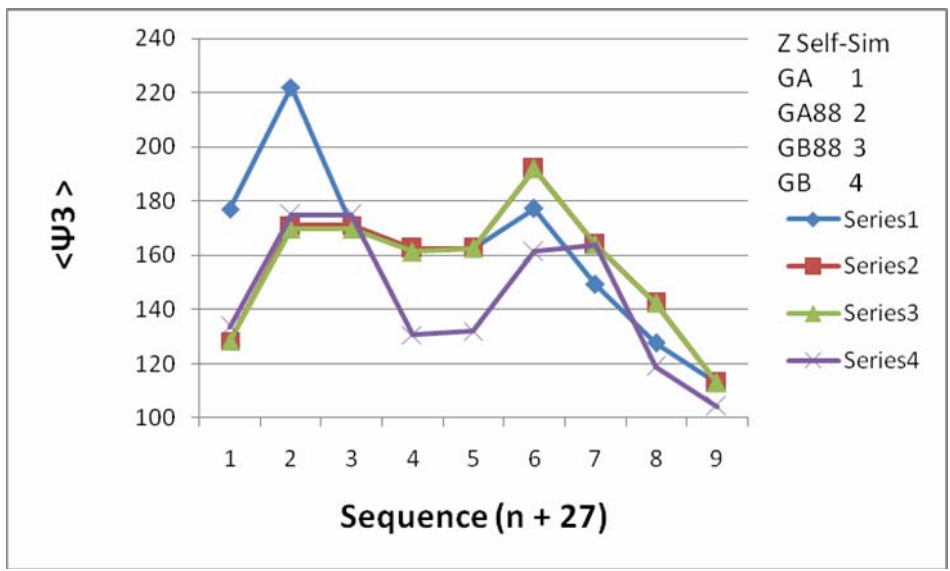

Fig. 2

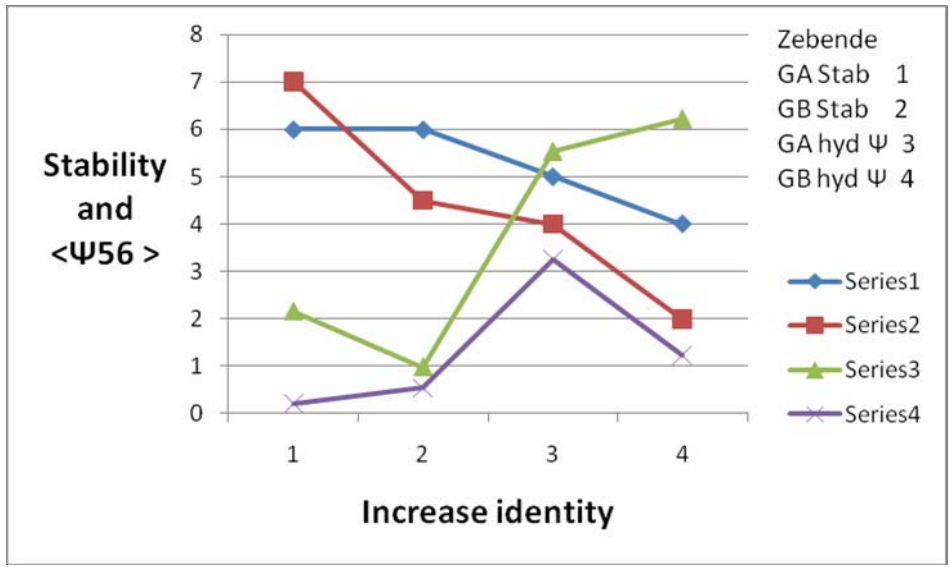

Fig. 3

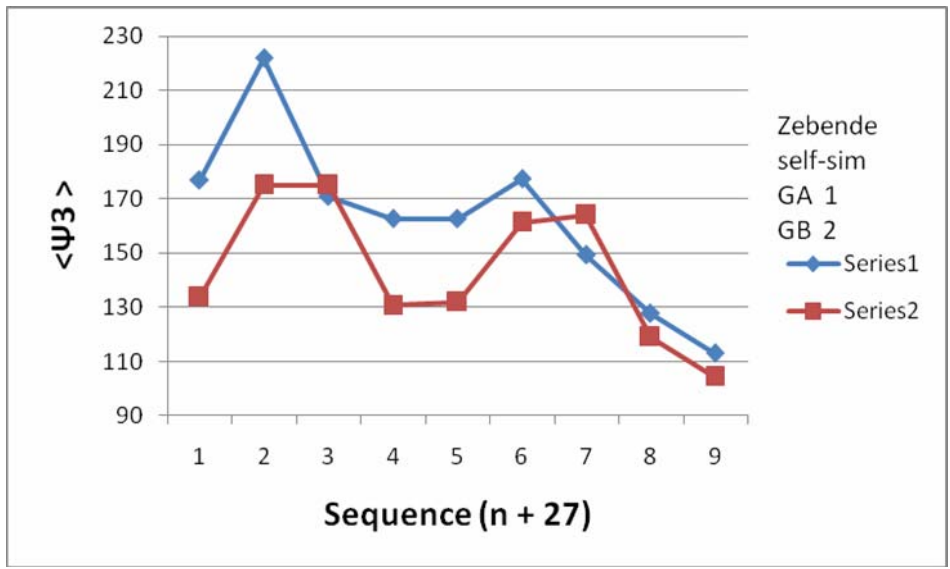

Fig. 4(a)

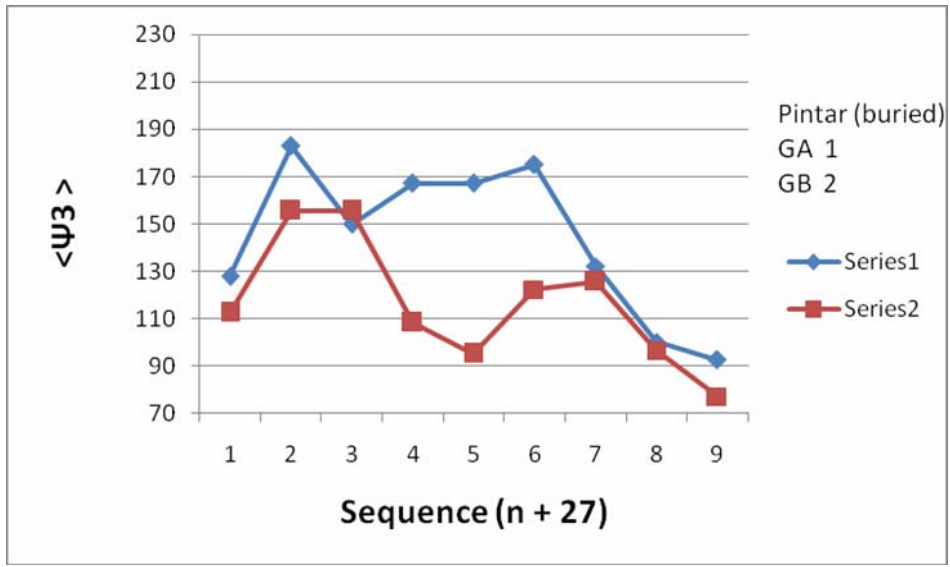

Fig. 4(b)

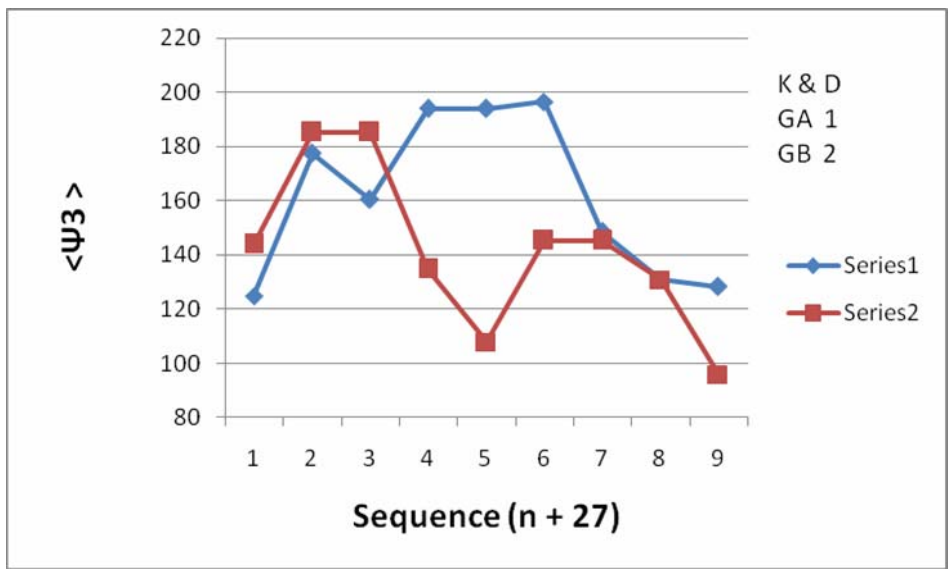

Fig. 4 (c)

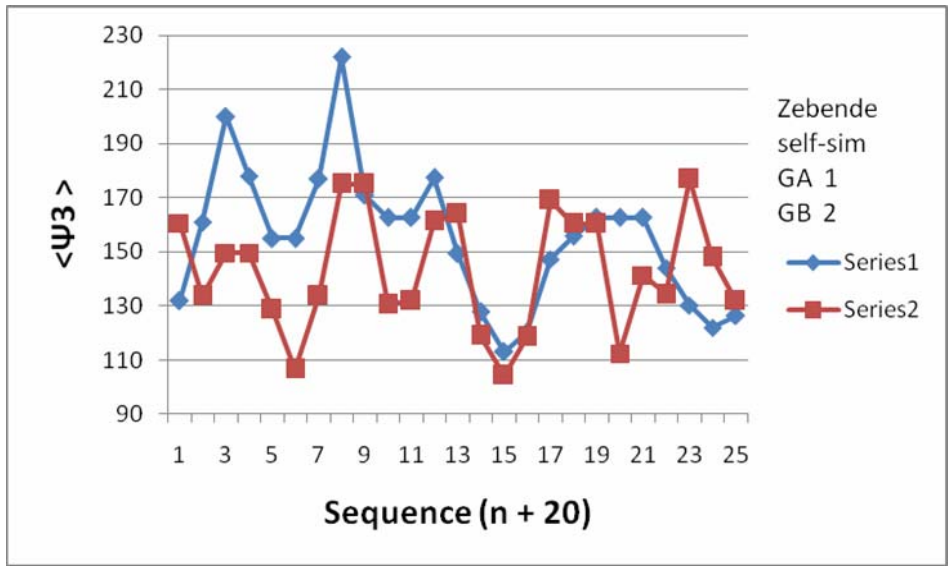

Fig. 5(a)

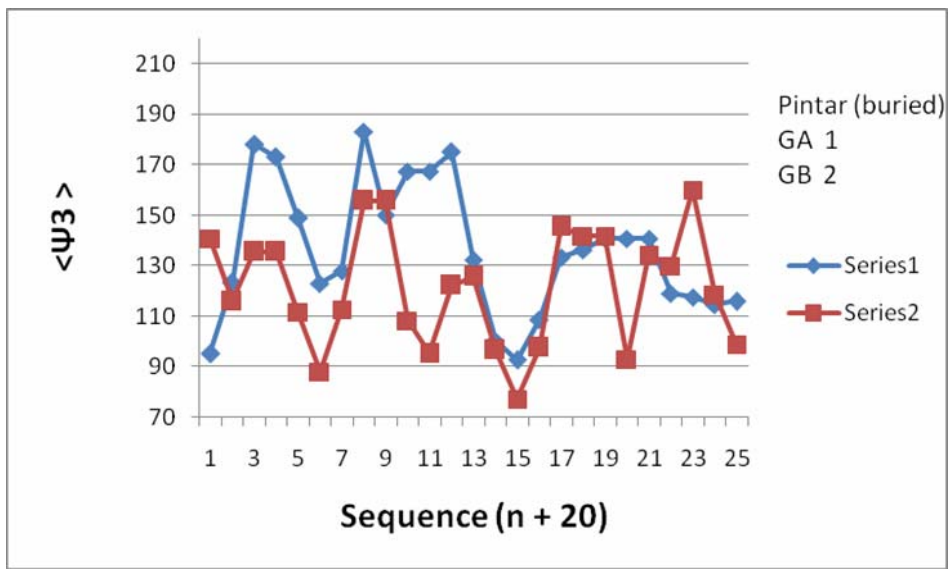

Fig. 5(b)

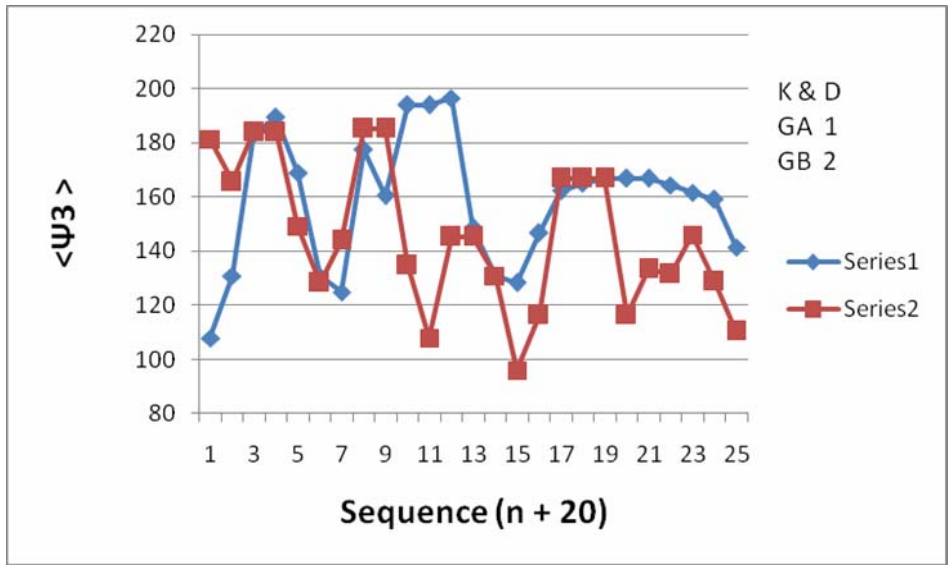

Fig. 5(c)

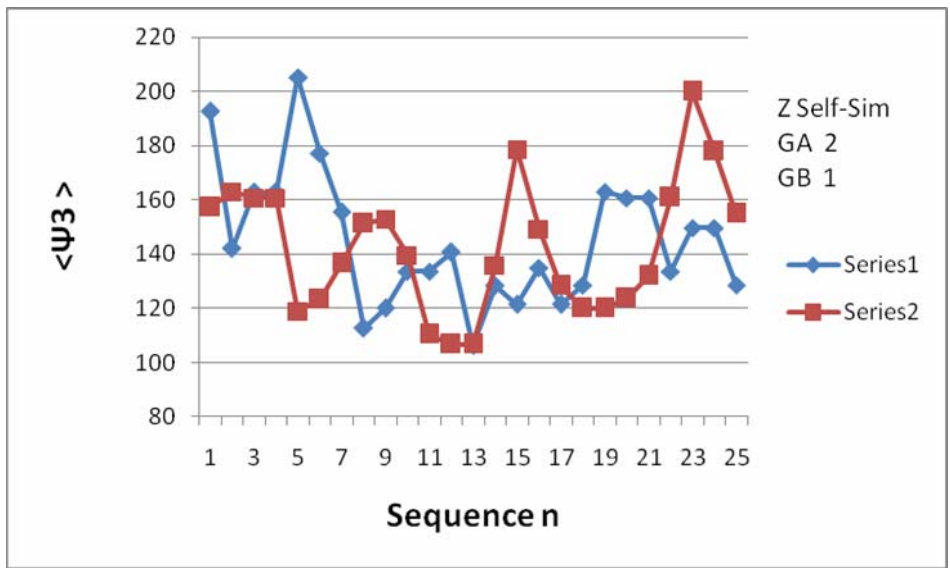

Fig. 6 (a)

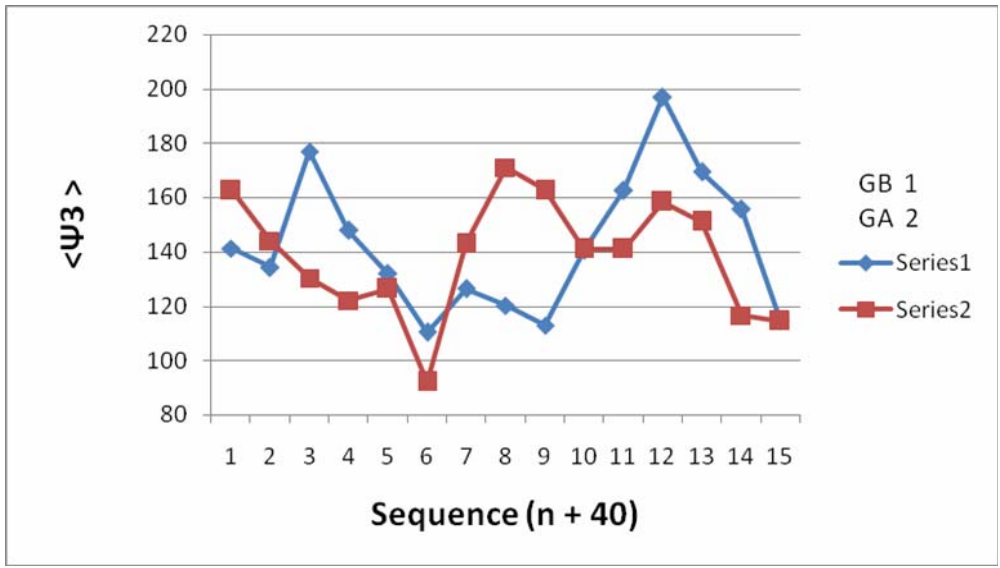

Fig. 6 (b)